# Electrically controlled superconductor-insulator transition and giant anomalous Hall effect in kagome metal $CsV_3Sb_5$ nanoflakes

Guolin Zheng[1, 2#], Cheng Tan[1#], Zheng Chen[2#], Maoyuan Wang[3], Xiangde Zhu[2], Sultan Albarakati[1], Meri Algarni[1], James Partridge[4], Lawrence Farrar[1], Jianhui Zhou[2*], Wei Ning[2*], Mingliang Tian[2,6], Michael S. Fuhrer[5], Lan Wang[1*]

[1]ARC Centre of Excellence in Future Low-Energy Electronics Technologies (FLEET), School of Science, RMIT University, Melbourne, VIC 3001, Australia.

[2]Anhui Province Key Laboratory of Condensed Matter Physics at Extreme Conditions, High Magnetic Field Laboratory, HFIPS, Chinese Academy of Sciences (CAS), Hefei 230031, Anhui, China.

[3]International Center for Quantum Materials, School of Physics, Peking University, Beijing 100871, China.

[4]School of Science, RMIT University, Melbourne, VIC 3001, Australia.

[5]ARC Centre of Excellence in Future Low-Energy Electronics Technologies (FLEET), Monash University, Melbourne, VIC 3800, Australia.

[6]Department of Physics, School of Physics and Materials Science, Anhui University, Hefei 230601, Anhui, China.

[#] These authors equally contributed to the paper.

[*] Corresponding authors. Correspondence and requests for materials should be addressed to J. Z. (jhzhou@hmfl.ac.cn), W. N. (email: ningwei@hmfl.ac.cn) and L. W. (email: lan.wang@rmit.edu.au).

# Abstract

The electronic correlations (e.g. unconventional superconductivity (SC), chiral charge order and nematic order) and giant anomalous Hall effect (AHE) in topological kagome metals $AV_3Sb_5$ (A= K, Rb, and Cs) have attracted great interest. Electrical control of those correlated electronic states and AHE allows us to resolve their own nature and origin and to discover new quantum phenomena. Here, we show that a protonic gate can largely modulate the effective disorders and carrier density in $CsV_3Sb_5$ nanoflakes, leading to significant modifications of SC, unusual charge density wave (CDW) and giant AHE. Notably, we observed a direct superconductor-insulator transition (SIT) driven by superconducting phase fluctuation due to the doping-enhanced disorders, in addition to a large suppression of CDW. Meanwhile, the carrier density modulation shifts the Fermi level across the CDW gap and gives rise to a nontrivial evolution of AHE, in line with the asymmetric density of states of CDW sub-bands near the saddle point. With the first-principles calculations, we suggest the extrinsic skew scattering of holes in the nearly flat bands with finite Berry curvature by multiple impurities accounts for the giant AHE. Our work uncovers a disorder-driven bosonic SIT, outlines a global picture of the giant AHE and reveals its correlation with the unconventional CDW in the $AV_3Sb_5$ family.

The layered kagome metals AV$_3$Sb$_5$ (A=K, Rb and Cs) that possess topological electron bands and geometrical frustration of vanadium lattices are of great interests [1-3]. This is in no small part due to the many quantum phenomena that they support including unconventional SC [4-11], novel nematic order [12], chiral charge density order [13-22], giant anomalous Hall effect (AHE) [23-25] as well as the interplay between two-gap SC and CDW in CsV$_3$Sb$_5$ [26]. The unique coexistence of electronic correlations and band topology in AV$_3$Sb$_5$ allows for investigating intriguing transitions of these correlated states, such as the superconductor-insulator transition (SIT), a protocol quantum phase transition (QPT) that is usually tuned by disorders, magnetic fields and electric gating [27,28]. Moreover, the origin of giant AHE in AV$_3$Sb$_5$ and its correlation with chiral CDW remain elusive [29,30], in spite of several recently proposed mechanisms including the extrinsic skew scattering of Dirac quasiparticles with frustrated magnetic sublattice [23], the orbital currents of novel chiral charge order [13] or the chiral flux phase in the CDW phase [31]. Thus the ability to tune the carrier density and the corresponding Fermi surfaces would play a vital role in understanding and manipulating these novel quantum states and further realizing exotic QPTs.

In this letter, we report significant modifications of SC, CDW phase and giant AHE in topological kagome metal CsV$_3$Sb$_5$ nanoflakes with protonic gates [32,33]. Remarkably, we observed a SIT associated with the localized Cooper pairs in thin CsV$_3$Sb$_5$ nanoflakes as the consequence of the doping-induced disorders, in addition to a dramatic suppression of CDW. In thicker CsV$_3$Sb$_5$ nanoflakes, however, the protonic gate mainly affects the carrier density (with the modulation up to $10^{22}\ cm^{-3}$) and the relevant Fermi surface topology changes from a hole pocket to an electron pocket. We find that the giant anomalous Hall conductivity (AHC) with a maximal amplitude exceeding $10^4\ \Omega^{-1}cm^{-1}$ is mainly pinned down to a narrow hole-carrier-density window around $p = (2.5 \pm 1.2) \times 10^{22}\ cm^{-3}$ at low temperatures. Meanwhile, the AHE exhibits a clear extrinsic-intrinsic transition as the Fermi level shifts across the CDW gap near the saddle point. The observed giant AHE can be ascribed to the extrinsic skew scattering of the holes in the flat bands with nonzero Berry curvature by V vacancies and magnetic field tilted paramagnetic (PM) impurities.

The layered material CsV$_3$Sb$_5$ has a hexagonal crystal structure with space group $P6/mmm$ (No. 191). As shown in the upper panel of Fig. 1(a), the V$_3$Sb layers are sandwiched by antimonene layers and Cs layers. X-ray diffraction (XRD) (see Fig.S1), reveals a sharp (001) diffraction peak, indicating a single crystal possessing (001) preferred orientation, in line with previous work [4]. The striking feature of CsV$_3$Sb$_5$ is that the V atoms form a 2D kagome network. This frustrated magnetic sublattice of V was expected to induce novel correlation effects, such as spin-liquid states [34,35]. The lower panel of Fig. 1(a) illustrates a schematic of the gating device. A CsV$_3$Sb$_5$ nanoflake is mounted on the solid proton conductor with an underlying Pt electrode to form a solid proton field effect transistor (SP-FET).

Figure 1(b) shows the temperature-dependent sheet resistance $R_s$ (resistance per layer) curves of device #5 with thickness of 21 nm at various gate voltages. A clear SC phase appears with the offset transition temperature $T_c^{offset}$ around 3 K in the absence of a protonic gate. Besides, a resistance anomaly as a characteristic of CDW near

80 K can be identified on $R_s - T$ curve (around 90 K in bulk, also in Fig. S3). Applying a protonic gate, both SC and CDW are gradually suppressed and $R_s$ fattens near the value of quantum resistance of electron pairs $R_Q = h/4e^2$ at $V_g = -11$ V. At $V_g < -11 V$, temperature-dependent $R_s$ gradually exhibits an upturn at low temperature region and it eventually reaches up to $10^4 R_Q$ at $V_g = -25$ V, clearly demonstrating a characteristic of an insulator as shown in Fig. 1(b). Thus, a direct SIT is observed for the first time in CsV3Sb5 nanoflakes through the protonic gating. It may result from the enhanced effective disorders during the proton intercalations in the thinner nanoflakes (~20 nm) with relatively lower carrier density [36]. Note that, in the thicker CsV3Sb5 nanoflakes (~100 nm), the protonic gating mainly affects the carrier density and leads to a slight modulation of SC transition temperature but not SIT, which is consistent with recent results [37-39].

SIT can be usually characterized by two distinct scenarios (bosonic and fermionic) according to the nature of insulating phase, while the finite-size scaling analysis could yield its critical exponents and further reveal the universality class of QPT. To get the critical exponents of SIT, we plot more than twenty sets of $R_s(T, B)$ curves and find that they can collapse onto a single function, as predicted for a 2D SIT (also in Sec. S4). The appearance of flatten resistance near $R_Q$ suggests the bosonic nature of SIT, in which the coherent Copper pairs in the SC phase are localized by disorders with loss of macroscopic phase coherence in the insulating phase [27,28]. The finite size scaling dependence of $R_s$ on T and a tuning parameter has the form $\rho(T, n_s) = \rho [T/T_0(n_s)]$ [40-42] with $T_0 \propto |n_s - n_c|^{\nu z}$ where $n_s$ is the charge density and $T_0$ is the scaling parameter which approaches to zero at $n_s = n_c$. $\nu$ is the correlation-length exponent and z is the temporal critical exponent. By extracting the exponent product $\nu z$ and plotting $lnT_0$ versus $ln|n_s - n_c|$ curve, we can obtain $\nu z = 1.85$ with an uncertainty of $\pm 0.14$ (see Fig. S4-1). The estimated exponent product (vz = 1.85) is close to that of magnetic-field tuned SIT in the hybrid system of SC indium islands deposed on 2D indium oxide thin film [43], which was also attributed to localization of persisting Cooper pairs. Note that the critical exponent here is distinct from those of 2D conventional models for SIT [28] such as classical percolation model (vz = 4/3), quantum percolation model (vz = 7/3), which probably results from the complexity of SC gap and multiple impurities [44,45].

Figure 1(d) shows the derivatives of the RT curves at various $V_g$. Interestingly, CDW transition temperature $T_{CDW} = 85$ K at $V_g = 0$ V gradually decreases to 73 K at $V_g = -6.5$ V. More importantly, at $V_g \leq -7$ V, we found this resistance anomaly totally disappeared on RT curve, indicating the disappearance of the CDW. This suppression of CDW is consistent with recent works for $CsV_3Sb_5$ under high pressures, possibly due to band reconstructions or Fermi level shift [4-8].

Let us turn to demonstrate the significant tune of the Fermi surface topology and AHE by the gate voltage in device #4 with thickness around 80 nm. Figure 2(a) shows the Hall traces of device #4 at various temperatures and selected gate voltages. At low magnetic fields, the Hall resistance $R_{xy}$ at $V_g = 6.4$ V exhibited a nonlinear behavior at 5 K. This antisymmetric "S"-shape $R_{xy}$ was attributed to field induced AHE in $KV_3Sb_5$ [23] and $CsV_3Sb_5$ [24]. At high fields, $R_{xy}$ exhibits an approximately linear field dependence associated with the ordinary Hall effect induced by the Lorentz force. When 6.4 V ≥ $V_g$ ≥ −2.7 V, the Hall traces in device #4 exhibit two distinct features. For each gating voltage, the temperature-dependent Hall effects demonstrate a sign reversal at the critical temperature $T^*$ (Lifshitz transition temperature) due to the temperature-induced Lifshitz transition. In addition, the Hall slope decreases gradually as the voltage is swept towards −2.7 V, indicating a gradual increase of the hole carrier density. At $V_g = -4.6$ V, however, the Fermi surface topology suddenly changes from a hole pocket to an electron pocket with a negative Hall slope (also in Sec. S5). In contrast to the hole pockets, the Hall traces in the electron pockets exhibit no sign reversal as the temperature is increased, as shown in the bottom right panel of Fig. 2(a) at $V_g = -6.1$ V, indicating a dramatic suppression of $T^*$ in the electron pocket. We further plot the gate-dependent carrier density at 5 K in device #4 in Fig. 2(b). At $V_g = -4.6$ V, the Fermi level is shifted across the electro-hole crossover point. Figure 2(c) shows the carrier-density (obtained at 5 K) dependent Lifshitz transition temperatures $T^*$ for bulk crystals and four nanodevices (devices #1-4). In hole pockets, a higher hole density may lead to a smaller $T^*$. However, $T^*$ approaches 0 K for the electron pockets, due to the sudden change of the Fermi surface topology.

We now discuss the gate-dependent AHE in CsV$_3$Sb$_5$. The total Hall resistivity $\rho_{xy}$ consists of two components [46]: $\rho_{xy} = \rho_{xy}^N + \rho_{xy}^A$, with $\rho_{xy}^N$ the normal Hall resistivity and $\rho_{xy}^A$ the anomalous Hall resistivity. In order to extract the AHE component, the Hall resistivity was linearly fitted at high field to subtract $\rho_{xy}^N$ (see Section S6). Figure 3(a) shows the gate-dependent anomalous Hall resistivity $\rho_{xy}^A$ of device #4 at 5 K under various gate voltages. The maximum $\rho_{xy}^A$ occurs at $V_g = 4.5$ V with an amplitude of 0.041 $\mu\Omega \cdot cm$ that is approximately eight-fold larger than the minimum $\rho_{xy}^A$ (0.0048 $\mu\Omega \cdot cm$) measured at $-4.6$ V. Interestingly, the AHE also exhibits a sign reversal at $V_g = -4.6$ V which is probably due to the sign change of the Berry curvature in different energy bands, as shown in Fig. 3(a). Figure 3(b) displays the non-monotonic variation of both the AHC $\sigma_{xy}^A = -\rho_{xy}^A/\left({\rho_{xy}^A}^2 + \rho_{xx}^2\right)$ and the anomalous Hall angle (AHA) $\theta = \left|\sigma_{xy}^A/\sigma_{xx}\right|$. The maximal AHC reaches $1.25 \times 10^4 \ \Omega^{-1}cm^{-1}$ with an AHA of 2.2% at 4.5 V. Moreover, the AHC (AHA) can be modulated by more than a decade in device #4, revealing the high tunability of the AHE in CsV$_3$Sb$_5$. Figure 3(c) shows two carrier density regions that exhibit large AHE for different devices. The first region is mainly pinned down in the hole pocket around $p = (2.5 \pm 1.2) \times 10^{22} \ cm^{-3}$ with the maximum AHC exceeding $10^4 \ \Omega^{-1}cm^{-1}$. Remarkably, another large but negative AHE appears in the electron pocket between $n = (3 \pm 0.6) \times 10^{22} \ cm^{-3}$ with the AHC around $5 \times 10^3 \ \Omega^{-1}cm^{-1}$. Shifting away from these two regions, however, the AHC either keeps a finite value or approaches zero for devices #1 and #2 (details in Section S7).

The scaling law between AHC and $\sigma_{xx}$ may assist in identifying the underlying mechanism of the AHE [46,47]. Figure 4(a) displays the scaling relations $\sigma_{xy}^A$ vs $\sigma_{xx}$ at various gate voltages and temperatures. In the high conductivity region ($\sigma_{xx}$ exceeds $5 \times 10^5 \ \Omega^{-1}cm^{-1}$), the AHE in device #4 can be well captured by a linear scaling relation $\sigma_{xy}^A \propto 0.14 \ \sigma_{xx}$ at 4.5 V, revealing that the skew-scattering mechanism may dominate the AHE [46-50]. The possible side jump contribution is also discussed in

Section S14. However, at $V_g = -4.6$ V, a finite AHC around $10^3$ $\Omega^{-1}cm^{-1}$ is approximately independent of the longitudinal conductivity $\sigma_{xx}$, implying that the intrinsic AHE from the Berry curvature becomes dominant. For other gating voltages, AHEs are likely linked to the mixing region. The gate-induced crossover between the extrinsic (at $V_g = 4.5$ V) and intrinsic regimes (at $V_g = -4.6$ V) reveals a strong dependence of AHE on the Fermi energy of CsV$_3$Sb$_5$.

To gain more insights into AHE, we performed the first-principles calculations of the band structure, the density of states (DOS) and the intrinsic AHC (in Figs. S8-S9). The calculated AHC over a broad energy region exhibits a maximum (~1500 $\Omega^{-1}cm^{-1}$) in the hole band [51], one order smaller than the maximum experimental value. This suggests that the intrinsic contribution from the Berry curvature of energy bands should not dominate the giant AHE in the hole pocket [13, 23, 24]. It has been known that the extrinsic skew scattering of AHE essentially originates from the asymmetric scattering of carriers by nonmagnetic/magnetic impurities [44]. Usually, there are three distinct scenarios to produce the extrinsic skew scattering and the resultant AHE [46]. By careful examination of each scenario, we could exclude the Kondo scattering and resonant skew scattering (Section S13). We found that the scenario associated with finite Berry curvature of energy bands and scattering by nonmagnetic/magnetic impurities probably accounts for extrinsic AHE in AV$_3$Sb$_5$ [46,52].

We further investigate the impact of charge doping on the band structure and AHE. Since the charge doping in CsV$_3$Sb$_5$ is orbitally selective, the hole (electron) doping can significantly shift the van Hove singularity (VHS1) upward (downward) with respect to the Fermi level within the rigid-band approximation, as shown in Fig. 4(b). In our pristine CsV$_3$Sb$_5$ single crystal, the Fermi level lies slightly above VHS1 near the M point (Section S3), with some nearly flat bands consist of $d_{xz,yz}$ and $d_{xy, x^2-y^2}$ orbitals of V atoms (Fig. S10) [53-55]. When T < $T_{cdw}$, a CDW gap opens near VHS1, splits the bands at VHS1 into two sub-bands and suppresses the DOS near the Fermi level, as shown in Fig. 4(c). Accordingly, the Fermi level in bulk CsV$_3$Sb$_5$ lies in the CDW gap [11,53] near the M point (red dashed arrow), exhibiting a large AHE. In exfoliated CsV$_3$Sb$_5$ nanoflakes, the Fermi level approaches the lower sub-band due to the increasing Cs vacancies and the AHC at $V_g = 0$ V reduces to about one third of the maximal value, i.e., $4500\ \Omega^{-1}cm^{-1}$. Applying a negative voltage will accordingly lower the Fermi level (details in Section S12) and generate a relatively large but negative AHE region (with AHC around $5000\ \Omega^{-1}cm^{-1}$) in the electron pockets. At $V_g > 0\ V$, however, the Fermi level will be shifted upward and back to the upper sub-band again (dashed red arrow), the giant AHE reappears at $4.5\ V$ in device #4. This giant AHE primarily comes from the skew scatterings of holes in the nearly flat bands at VHS1 with nonzero Berry curvature by the V vacancies and/or PM impurities [56, 57]. The large discrepancy of AHE in the electron and hole pockets is consistent with asymmetric distribution of DOS in the CDW bands near the M points [11]. Note that the intrinsic AHC at $V_g = -4.6$ V near the electron-hole crossover point is mainly ascribed to the large suppression of the DOS at the middle of the CDW gap. This intrinsic AHE may originate from the recently suggestive chiral charge order forming from the electronic states near the saddle point [58-61].

In summary, by use of protonic gates, we observed a bosonic SIT associated with localized Cooper pairs driven by the effective disorders in CsV$_3$Sb$_5$ nanoflakes. The large modification of the carrier density induces a clear extrinsic-to-intrinsic transition of AHE and leads to a nontrivial evolution of AHE that is in line with the asymmetric DOS of CDW sub-bands around the VHS1. The giant AHE in AV$_3$Sb$_5$ can be attributed to the intense extrinsic skew scattering of holes in the nearly flat bands with finite intrinsic AHE in the CDW phase at the saddle points by multiple impurities. This significant and electrically controlled SIT and AHE in AV$_3$Sb$_5$ should inspire much investigations of the relevant intriguing physics and promising energy-saving nanoelectronic devices.

## Acknowledgements

The authors thank Yaomin Dai, Harrison LaBollita, Haiwen Liu, Kosuke Nakayama, Qian Niu, Takafumi Sato, Cong Xiao, Zhenyu Wang, Huan Yang, Jianjun Ying and Li Yu for insightful discussions. This research was supported by the Australian Research Council Centre of Excellence in Future Low-Energy Electronics Technologies (Project No. CE170100039). This work was also partially supported by the High Magnetic Field Laboratory of Anhui Province.

of Sb atoms (see Fig. S10) remain almost unchanged during the CDW transition [57] and does not possess finite Berry curvature. Thus, they are irrelevant to the AHE. Similarly, the Dirac bands that open a small CDW gap [57] merely make minor contribution to AHE (also see Section S13).

**Figure caption**

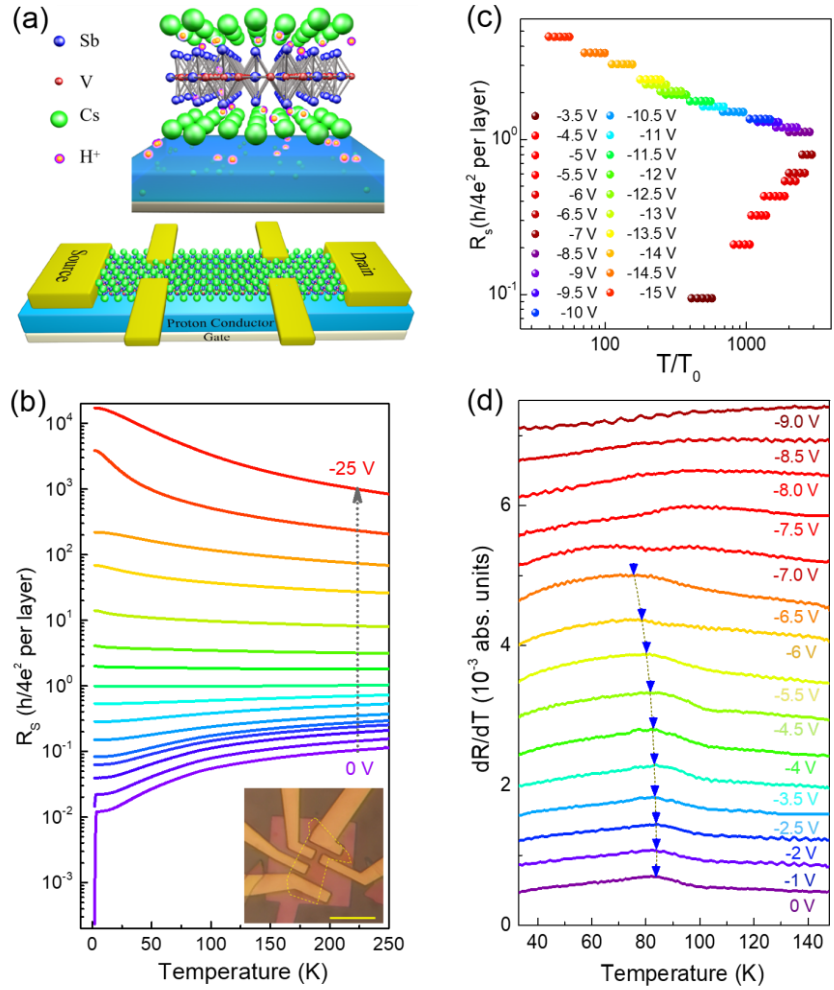

FIG. 1. Temperature-dependent longitudinal resistance curves under various gating voltages in device #5 with thickness 21 nm. (a) Schematic of proton gating on $CsV_3Sb_5$ nanoflakes (upper) and Hall-bar device (lower). (b) Temperature dependence of resistance per layer at various gate voltages in device #5 exhibits a clear separatrix near the quantum resistance of pairs $R_Q = h/4e^2$, indicating a direct SIT. Inset: Optical image of device #5 on a solid proton conductor. Scale bar: 10 μm. (c) Multiple sets of $R_s(T, B)$ curves can collapse onto a single function, as predicted for a 2D SIT. (d) shows the derivatives of the resistance curves $R_{xx}(T)$ under different voltages. As the voltage changes from 0 V to −6.5 V, the CDW transition temperature $T_{cdw}$ gradually decreases from 85 K to 73 K at $V_g = -6.5$ V. CDW is largely suppressed when the gate voltage exceeds −7.0 V.

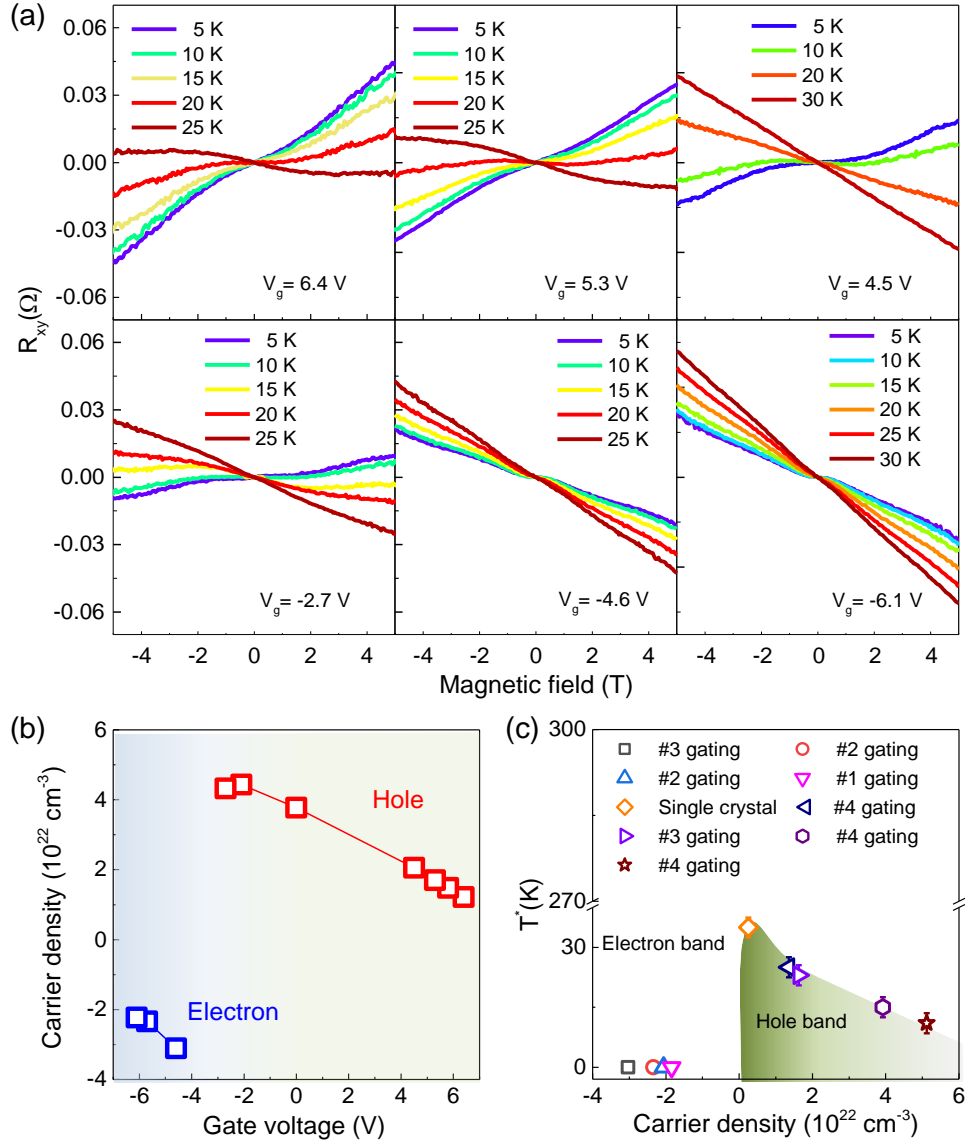

FIG. 2. Gate-tuned Hall resistance and carrier density dependent band topology. (a) Temperature-dependent Hall effect in device #4 under different gating voltages. (b) Gate-dependent carrier density in device #4 at 5 K. Sweeping the gate voltage from 6.4 V to −6.1 V, the band structure evolves from a hole band to an electron band in the low temperature region. (c) Carrier density dependent Lifshitz transition points $T^*$ in different samples.

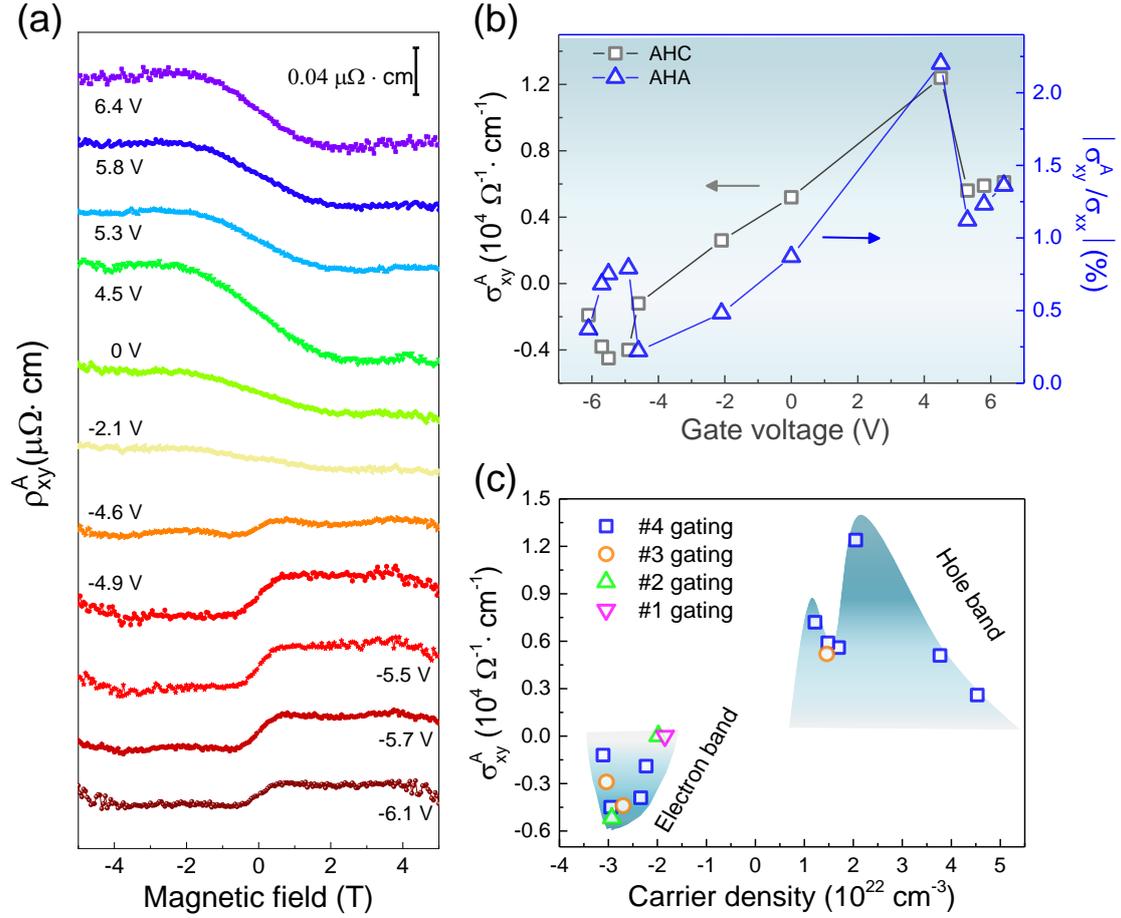

FIG. 3. Gate-tuned giant anomalous Hall effects in device #4. (a) Gate-dependent anomalous Hall effect at 5 K after subtraction of the linear Hall background in the high field regions (ordinary Hall part). (b) Gate-dependent AHC and anomalous Hall angles (AHA). (c) Carrier density dependent AHC in different devices #1, #2, #3 and #4. The maximum AHC occurred with a hole carrier density of approximately $2 \times 10^{22}\ cm^{-3}$.

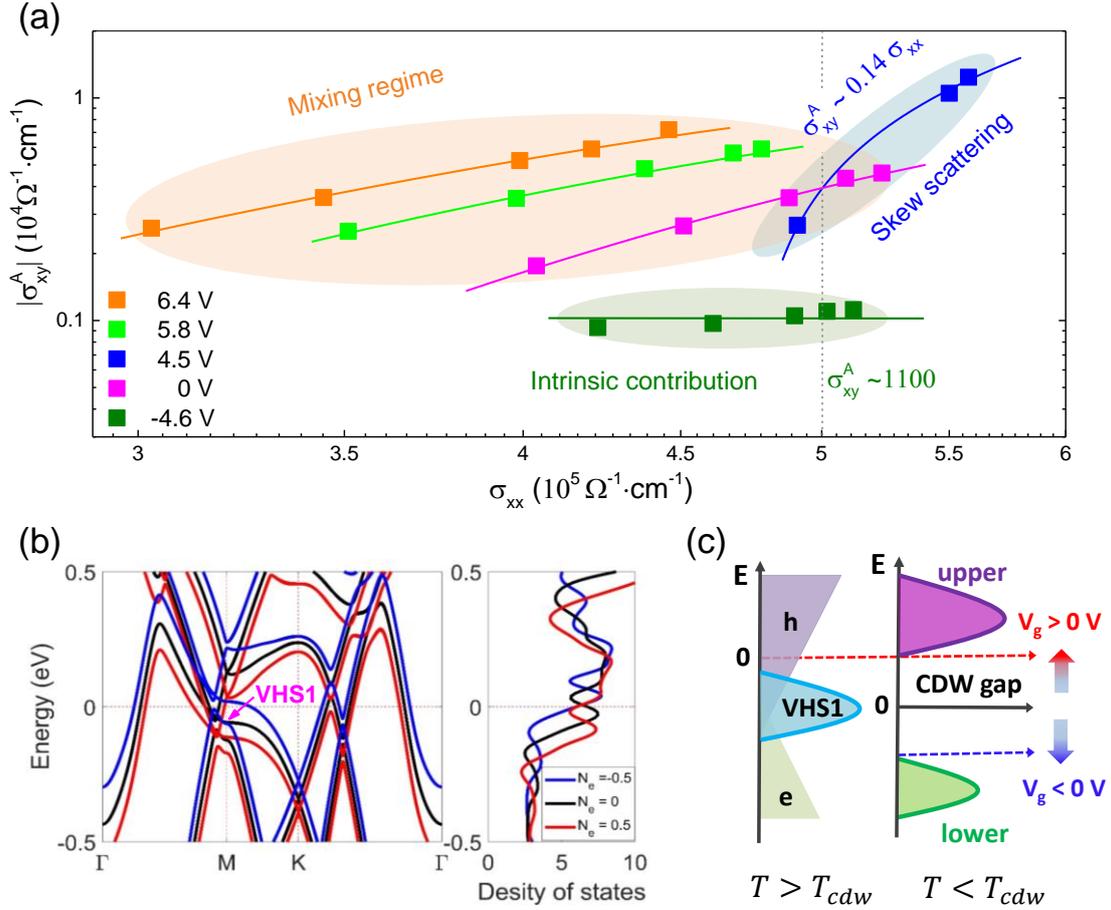

FIG. 4. Scaling relation, band structure and the density of states in gated CsV$_3$Sb$_5$. (a) The scaling relation of AHC against the longitudinal conductivity in device #4. In the high conductivity region (above $5 \times 10^5 \ \Omega^{-1} cm^{-1}$), the AHE is dominated by the skew scattering (at 4.5 V, hole band). At $-4.6$ V (electron band), the AHE is dominated by the intrinsic Berry curvature. (b) Band structures of the paramagnetic phase with different doping levels in CsV$_3$Sb$_5$. $Ne$ refers the charge number in each primitive cell. (c) Illustration of the evolution of Fermi energy under different gate voltages. Red dashed arrow shows the probable Fermi level in bulk crystal (slightly above VHS1). Cs vacancies in CsV$_3$Sb$_5$ nanoflakes will significantly lower the Fermi level (black arrow). Applying a negative (positive) voltage will shift the Fermi level downward (upward). The giant AHE occurs when the Fermi level approaches the upper sub-band with relatively large DOS.